\documentclass{JHEP3}

\usepackage{epsfig}
\usepackage{amsmath}
\usepackage{amssymb}

\def\e{{\,\rm e}\,}
\def\d{{\rm d}}
\def\i{{\rm i}}
\def\K{K}
\def\C{C}
\def\D{{\cal D}}
\def\Tau{{\cal T}}

\newcommand{\rf}[1]{(\ref{#1})}
\newcommand{\eq}[1]{Eq.~(\ref{#1})}
\def\be{\begin{equation}}
\def\ee{\end{equation}}
\def\beq{\begin{equation}}
\def\eeq{\end{equation}}
\def\bea{\begin{eqnarray}}
\def\eea{\end{eqnarray}}

\newcommand{\non}{\nonumber \\*}
\newcommand{\ie}{{i.e.}\ }

\newcommand{\pint}{\int\hspace{-1.2em}\not\hspace{.75em}}
\def\la{\lesssim}
\def\ga{\gtrsim}

\preprint{ITEP--TH--17/10}

\title{Semiclassical Regge trajectories of 
 noncritical string \\ and large-$N$ QCD}

\author{Yuri Makeenko%
\thanks{Also at the Institute for Advanced Cycling,
Blegdamsvej 19, 2100 Copenhagen \O, Denmark} 
\\Institute of Theoretical and Experimental Physics\\
B.~Cheremushkinskaya 25, 117218 Moscow, Russia \\
E-mail: \email{makeenko@itep.ru}
}

\author{Poul Olesen$^*$ \\ 
The Niels Bohr International Academy,
The Niels Bohr Institute\\
Blegdamsvej 17, 2100 Copenhagen \O, Denmark\\
E-mail: \email{polesen@nbi.dk} 
}

\date{\today}
\abstract{
By properly treating the path integral over the boundary value
of the Liouville field (associated with reparametrizations
of the boundary contour) in open string theory, we derive 
consistent off-shell scattering amplitudes in $d=26$ dimensions.
In $d<26$ we consider a recently proposed boundary ansatz which
reproduces a semiclassical correction to the classical string 
(known as  the L\"uscher term) and obtain in the semiclassical approximation  
a linear Regge trajectory with the intercept $(d-2)/24$. 
We associate it with the quark-antiquark Regge trajectory in
large-$N$ QCD and explain why it dominates over perturbative
QCD when $t>-{\rm few~GeV}^2$.}


\dedicated{Dedicated to the 40th Anniversary of QFT/string correspondence} 

\begin{document}

\maketitle
\setcounter{page}{2}

\section{Introduction}

It is commonly believed that the  Nambu--Goto and Polyakov strings
are equivalent. At the classical level this stems from the 
fact that the intrinsic metric coincides with the induced one.
At the quantum level the tree amplitude of scattering of scalar particles
with momenta $\Delta p_m$ reads for the Polyakov formulation of 
an open bosonic string in the critical dimension $d=26$,
where the Liouville field decouples in the bulk:
\be
A\left( \{ p_m \} \right) = \int \D \varphi(s) \int \prod _{m} \d s_m
\, \e^{\varphi(s_m)/2-\pi \alpha'\Delta p_m^2 G(s_m,s_m)} 
\prod_{j\neq m} |s_j-s_m|^{\alpha' \Delta p_j\cdot \Delta p_m} \,,
\label{ampl1}
\ee 
where $-\infty<s_m<s_{m+1}<+\infty$ 
are the Koba--Nielsen variables. 
The path integration
is over the boundary value $\varphi(s)$ of the Liouville field 
which remains as a dynamical variable.
It is included in invariant vertex operators, which explains its 
appearance on the right-hand side of \eq{ampl1}.

A very subtle matter with \eq{ampl1} is the singularity in the Green function 
at coinciding arguments,
which has to be regularized in an invariant way as~\cite{Pol87}
\be
 G(s_m,s_m) \longrightarrow  G_\varepsilon (s_m,s_m) 
=\frac1\pi \ln \frac 1\varepsilon + \frac{1}{2\pi} \varphi(s_m)\,,
\label{Geps0}
\ee 
where the appearance of the boundary value $\varphi(s_m)$ of the Liouville 
field is required for an invariant regularization.
There is now a precise mutual cancellation
of $\varphi(s_m)$ in the exponent on
the right-hand side of \eq{ampl1} when $\alpha' \Delta p_m^2=1$, 
i.e.\ for a tachyon.
Therefore, the integrand does not depend on  $\varphi(s)$ for
the tachyonic scalars or massless vectors, so
the path integration over the boundary value  $\varphi(s)$ of the Liouville 
field decouples and the standard on-shell Koba--Nielsen amplitudes 
are reproduced~\cite{Pol87,ADN86} for the Polyakov formulation. 

In this Paper we shall {\em not}\/ impose the on-shell condition and
study the amplitude~\rf{ampl1} by evaluating  the path integral
over $\varphi(s)$, utilizing the technique which has been recently
developed in Refs.~\cite{MO08,MO09,BM09,MO10}. 
By properly treating the path integral in \eq{ampl1}, we remarkably
obtain consistent off-shell scattering amplitudes similar to the ones known
in the literature as the Lovelace choice~\cite{Lov70,DiV92}.
In $d<26$ we consider a recently proposed ansatz~\cite{MO10}
for the disk amplitude with the Dirichlet boundary conditions,
that reproduces  a semiclassical correction to the classical
string (also known as the L\"uscher term). By evaluating the path
integral in the semiclassical approximation, we obtain a linear Regge 
trajectory with the intercept $(d-2)/24$ which remarkably coincides with 
the well-known result~\cite{Arv83,p85} derived from the spectrum 
of the Nambu--Goto string.  
We associate it with the quark-antiquark Regge trajectory in
large-$N$ QCD and explain why it dominates over perturbative
QCD in the Regge kinematical regime, when $s\gg -t$ and $t>-{\rm few~GeV}^2$.

\section{Discretized Green function}

In order to calculate the path integral over $\varphi(s)$ in \eq{ampl1},
we approximate it by an $N$-dimensional integral, discretizing 
the real axis by a set of points $s_i$'s.
Strictly speaking, it 
is better to discretize an angular variable
\be
\sigma=-2\; \hbox{arccot}\; s, 
\label{angular}
\ee
which takes the values in the finite interval $[0,2\pi)$ and whose
discretization can be chosen equidistant, but $\varphi(s)$
vanishes as $s\to \pm\infty$, so a proper discretization of $s$
also works.  

Having introduced a discretization of $s$, we simultaneously regularize
the Green function $G(s,s')$. 
Let $s$ lies on the $i$-th interval ($s_{i-1}<s\leq s_i$) and 
$s'\geq s$ may then lie either on the $i$-th or $i+1$-th interval 
($s \leq s' < s_{i+1}$) or outside. 
The regularized Green function reads
\be
\widetilde G(s,s') =\left\{
\begin{array}{ll}
\displaystyle{-\frac 1\pi}\ln (s'-s)
\qquad &\hbox{for}\quad s'\geq s_{i+1}\\[2mm]
\displaystyle{-\frac 1\pi \ln\frac{(s'-s_{i-1})(s_{i+1}-s)}{(s_{i+1}-s_{i-1})} }
\qquad &\hbox{for}\quad s\leq s'< s_{i+1} 
\end{array}
\right. 
\label{tildeG}
\ee 
which possesses the projective co-invariance and the continuity at $s'=s_{i+1}$.
The construction for $s\geq s'$ is analogous since $\widetilde G(s,s')$
is symmetric in $s$ and $s'$.

In particular, we have from \eq{tildeG}
\be
\widetilde G(s,s)=\displaystyle{\frac 1\pi 
\ln \frac{(s_{i+1}-s_{i-1})}{(s-s_{i-1})(s_{i+1}-s)} }
\label{tildeG0}
\ee
which replaces \eq{Geps0}. 

The appearance of $s_i$ in the above formulas is in fact related
to the discretization of the boundary value of the Liouville field.
Comparing Eqs.~\rf{Geps0}  and \rf{tildeG0}, we arrive at 
\be
\e^{\varphi(s)/2} \propto 
\frac{(s_{i+1}-s_{i-1})}{(s_{i+1}-s)(s-s_{i-1})}\,,\qquad  s_{i-1}<s \leq{s_i}
\label{discLiou}
\ee
which is in the spirit of the formula 
\be
\e^{\varphi(s)/2}=\sqrt{\dot x^2(s)} 
\label{112}
\ee
for the induced boundary metric.

For $s=s_i$ we have from \eq{discLiou}
\be
\e^{\varphi(s_i)/2} \propto 
\frac{(s_{i+1}-s_{i-1})}{(s_{i+1}-s_i)(s_i-s_{i-1})}
\label{discLiousi}
\ee 
providing the discretization of $\varphi(s_i)$ is through \eq{Geps0}. 

The presence of the ratio on
the right-hand side of \eq{discLiou} provides 
the correct formula
\be
\varphi(s) \to  \varphi\left( \frac{a s+b}{c s+d}\right)
=\varphi(s)+2 \ln (c s+d)^2
\label{protra}
\ee
for the projective transformation exactly at finite discretization,
which was a guiding principle of Ref.~\cite{MO09} 
for constructing the discretization.

Equation~\rf{protra} follows from the general law~\cite{Pol87}
of transformation of the boundary Liouville field $\varphi(s)$
under the reparametrization $s\to t(s)$ ($\d t/\d s\geq 0$):
\be
\varphi(s) \to \varphi(t(s))=\varphi(s)-2 \ln \frac{\d t(s)}{\d s}.
\label{115}
\ee
We see from this formula that the path integration over $\varphi(s)$
can be represented as a path integration over $t(s)$, \ie over
reparametrizations. The only difference resides in one mode of $\varphi(s)$,
associated with a scale, which can be ignored as is discussed in
Sect.~\ref{s:4} below.

\section{Projective-invariant scattering amplitude\label{s:3}}

Equations~\rf{tildeG0}, \rf{discLiou} fix the additional factor 
in \eq{ampl1} to be
\be
\e^{\varphi(t_m)/2-\pi \alpha ' \Delta p_m^2 G(t_m,t_m) } =
\left[ \frac{(s_{K_m+1}-s_{K_m-1})}
{(s_{K_m+1}-t_{m})(t_{m}-s_{K_m-1})}\right]^{-\alpha'  \Delta p_m^2+1}
\label{di1}
\ee
with $t_m \equiv s_{K_m}$, which has to be integrated over $s_i$'s at
intermediate points. This can be done using 
the following discretization of the measure:
\be
\D_{\rm diff} s = \prod_{i} \d s_i 
\left[ \frac{(s_{i+1}-s_{i-1})}{(s_{i+1}-s_{i})(s_{i}-s_{i-1})} \right],
\label{altmeasure}
\ee 
that supersedes the one
\be
\D_{\rm diff} s = \prod_{i} \d s_i 
\frac{1}{(s_{i}-s_{i-1})} 
\label{PRDmeasure}
\ee 
proposed in \cite{MO09}.
How to regularize integrals with 
the measure~\rf{altmeasure} is discussed in Appendix~\ref{appC}.
Both measures~\rf{altmeasure} and \rf{PRDmeasure} possess
the projective invariance at finite discretization.
There should be no difference between these
two discretizations for smooth trajectories but 
typical trajectories in the path integral over reparametrizations 
are  discontinuous~\cite{BM09}, so this may be not the case.
Actually, using the discretization~\rf{altmeasure}, we shall obtain 
a linear Regge trajectory with the unit intercept instead
of the one with zero intercept in Ref.~\cite{MO09}.

The integration over $s_i$'s at the intermediate points can be 
repeatedly done using the formula
\be
\int_{s_1}^{s_3} \d s_2 
\left[\frac{s_{42}}{s_{43} s_{32}}\right]\left\{
\left[\frac{s_{31}}{s_{32} s_{21}}\right] 
\left[\frac{s_{20}}{s_{21} s_{10}}  \right] +c\frac{1}{s_{21}^2} \right\}
\propto 
\left[\frac{s_{41}}{s_{43} s_{31}}  \right]
\left[\frac{s_{30}}{s_{31} s_{10}}  \right]+(c+1)\frac{1}{s_{31}^2} 
\label{altconvolution}
\ee
with $s_{ij}\equiv s_i-s_j$,
which is valid modulo an (infinite) factor for an analytic 
regularization of the  type of the $\zeta$-function, as is discussed 
in Appendix~\ref{appC}.

Integrating over $s_i$ at all intermediate points except two for
each interval: $v_m,u_{m-1}\in \left(t_{m-1},t_{m}\right)$ with $v_m>u_{m-1}$, 
we arrive at the amplitude
\begin{equation}
A\left(\{\Delta p_m \}\right)
= \int \prod_m \d t_m \,{\cal K}\left(\{t_m \}\right)
\prod_{j\neq m} \left(t_j-t_m \right)^{\alpha' \Delta p_j\cdot \Delta p_m}
 \,,
\label{amplitudep22}
\end{equation}
where
\bea
 \lefteqn{{\cal K}\left(\{t_m \}\right)} \non &&=
\prod_m\int_{t_{m-1}}^{t_{m}} \d v_m \int_{t_{m-1}}^{v_{m}} \d u_{m-1}
\left[ \frac{(u_{m}-v_{m})}
{(u_{m}-t_{m})(t_{m}-v_{m})}\right]^{-\alpha' \Delta p_m^2+1} 
D \left( t_m, v_m; u_{m-1}, t_{m-1}  \right) \non &&
\label{calKD} 
\eea
with 
\begin{eqnarray}
D \left( t_m, v_m; u_{m-1}, t_{m-1} \right)&
=&\left[\frac{(t_{m}-u_{m-1})}{(t_{m}-v_{m})(v_{m}-u_{m-1})} \right] 
\left[\frac{(v_{m}-t_{m-1})}{(v_{m}-u_{m-1})(u_{m-1}-t_{m-1})} \right] \non
&& + c \frac{1}{(v_{m}-u_{m-1})^2}\,,
\end{eqnarray}
which results from the path integral over reparametrizations as is shown
in Appendix~\ref{appC}.
It is obviously projective-invariant and most probably dual.

The simplest case is that of on-shell tachyons with $\alpha' \Delta p_m^2=1$.
Integrating over $v_m$'s and $u_{m-1}$'s by the formula
\be
{\cal K}\left(\{t_m \}\right)=\prod_m \int_{t_{m-1}}^{t_m} \d v_m 
\int_{t_{m-1}}^{v_m} \d u_{m-1}\, D\left(t_m,v_m;u_{m-1},t_{m-1}  \right)  
\stackrel{\delta \ll1} \propto 1
\,,
\ee
we then get the usual Koba--Nielsen amplitude with $\alpha(0)=1$.

For the off-shall case with arbitrary $\Delta p_m^2$ we find explicitly
for the 4-point amplitude, fixing projective invariance by
setting $t_1=0$, $t_2=x$, $t_3=1$, $t_4=\infty$ in the usual way:
\be
A\left(\{\Delta p_m \}\right)
= \int_0^1 \d x \,x^{-\alpha' s -\alpha' \Delta p_1^2 -\alpha' \Delta p_2^2}
(1-x)^{-\alpha' t -\alpha' \Delta p_2^2 -\alpha' \Delta p_3^2} \,{\cal K}(x)\,.
\label{amplitudep33}
\ee

The Regge asymptote of the amplitude~\rf{amplitudep33} can be analyzed by
the standard change of the variable
\be
x=1+\frac y {\alpha's }\,,\qquad 0<y<-\alpha's \,.
\ee
The values of $y$, which are essential in \rf{amplitudep33}, are $\sim 1$
because of the factor
\be
x^{-\alpha' s}=\left(1+\frac y {\alpha's }\right)^{-\alpha's }
\stackrel{-\alpha's \gg1}=\e^{-y}\,,
\ee
so that $(1-x)\sim 1/\alpha's$ are essential. 

Noting that an additional $s$-dependent factor in the 
amplitude~\rf{amplitudep33} comes from the integrals over $v_3$
and $u_2$, we have 
\bea
{\cal K}(x) &\propto& 
\int_x^1 \d v_3 \int_x^{v_3} \d u_2\,(1-v_3)^{\alpha' \Delta p_3^2-1} 
D\left( 1,v_3; u_2,x \right)
(u_2-x)^{\alpha' \Delta p_2^2-1}\non & \propto&
(1-x)^{\alpha' \Delta p_2^2+\alpha' \Delta p_3^2-2}.
\label{A14p}
\eea 
We thus find for the Regge asymptote of \rf{amplitudep33} 
\bea
A\left(\{\Delta p_m \}\right)&\stackrel{-\alpha's \gg1} \propto &
\int_0^1 \d x \,x^{-\alpha' s}
(1-x)^{-\alpha' t -\alpha' \Delta p_2^2-\alpha' \Delta p_3^2  } 
(1-x)^{\alpha' \Delta p_2^2+\alpha' \Delta p_3^2-2}\non
&=&\int_0^1 \d x \,x^{-\alpha' s }
(1-x)^{-\alpha' t -2},
\label{A15p}
\eea
resulting in the linear Regge trajectory 
\be
\alpha (t)=1+\alpha' t \,.
\ee
It is clear from this derivation that the same result for the Regge 
asymptote can be obtained directly from the path integral~\rf{ampl1},
employing the discretization~\rf{di1}, \rf{altmeasure}, without the
integration over the intermediate points.

Let us stress that the amplitude \rf{amplitudep33} is
not precisely the Lovelace amplitude~\cite{Lov70,DiV92}, but has the
same momentum-dependent part. The only difference resides in the
measure for the integration over the Koba--Nielsen variables.
The amplitude \rf{amplitudep33} is consistent off-shell in $d=26$
and can be viewed as
a straightforward off-shell extension of the Koba--Nielsen amplitudes, 
which possesses the projective invariance. 

\section{Wilson-loop/scattering-amplitude duality\label{s:4}}

The above scattering amplitude can be inferred from the 
reparametrization-invariant disk amplitude
\begin{equation}
W\left[x(\cdot)\right]=\int \D_{\rm diff} t \,\e^{-K S[x(t)]}\,,
\label{ansatz}
\end{equation} 
where $\K=1/2\pi\alpha'$ is the string tension and
\be
 S[x(t)]=\frac 1{4\pi}
\int\nolimits_{-\infty}^{+\infty} \d s_1 
\int\nolimits_{-\infty}^{+\infty} \d s_2 \,\frac{ \left[ 
 x(t(s_1))- x(t(s_2))\right]^2}{(s_1-s_2)^2} 
\label{Di}
\ee
is the Douglas integral~\cite{Dou31}, whose minimum with respect to
reparametrizations equals the minimal area.
Equation~\rf{ansatz} was proposed by Polyakov~\cite{Pol97} 
(see also Ref.~\cite{Ryc02}) as an ansatz for Wilson loops in large-$N$ QCD.

The derivation of the scattering amplitudes from the ansatz~\rf{ansatz}
makes use of the equivalence of the path integrals over the boundary
value $\varphi(s)$ of the Liouville field and the reparametrizations $t(s)$ of 
the boundary contour. They are the same modulo a scale factor which has to be
fixed to provide the correct value $L$ of the length of the boundary contour.
The precise equivalence of the two measures for the path integrations reads
\be
\int \D \varphi(s)\,\delta\left( \frac 1L \int \d s\,\e^{\varphi(s)/2}-1 \right)
\cdots = \int \D_{\rm diff} t \,\cdots \,.
\label{twomeasures}
\ee  
We can rephrase this equivalence by saying that the boundary metric
$\exp(\varphi/2)$ has to be equal to the induced metric~\rf{112}
modulo reparametrizations.

In our previous works~\cite{MO08,MO09} it was shown that the functional
Fourier transformation of \eq{ansatz}
\begin{equation}
A[p(\cdot)]\equiv\int \D x\,\e^{\i \int p \cdot \dot x}\,W[x(\cdot)],
\label{msWL}
\end{equation}
which defines a momentum-space disk amplitude, looks exactly like
the right-hand side of \eq{ansatz} with $x(t)$ substituted by
the ``most important'' trajectory
\be
x_*(t)= \frac{1}\K p(t).
\label{1st}
\ee 
For piece-wise constant $p(t)$ this momentum-space disk amplitude,
integrated over repara\-metrizations $s(t)$ obeying $s(t_m)=t_m$,
is equal (modulo an infinite factor) to  
the scattering amplitude~\rf{amplitudep22}.

The fact that the position-space and momentum-space disk amplitudes
coincide, if \eq{1st} is satisfied,
is in the spirit of the remarkable Wilson-loop/scattering-amplitude
duality, recently discovered for the ${\cal N}=4$ super Yang--Mills
\cite{AM07a,DSK07} (for a review see Ref.~\cite{AR08}).
As is shown in Refs.~\cite{MO08,MO09}, it is applicable to QCD, but
only in the Regge kinematical regime of meson scattering for not too
large values of $-t$.
Equation~\rf{1st} is the same as Eq.~(5) of \cite{MO09}, where it was 
discussed in detail. For  piece-wise constant $p(t)$, $x_*(t)$ is also a 
step function which has (if not regularized) discontinuities at
$t(s_m)=t_m$, that have to be smeared by a regularization. 


It is evident that a regularization of the Green function $G$, 
like the one displayed in \eq{Geps0}, can be moved from $G$ to the momentum 
loop $p(t)$. Let us take%
\footnote{This regularization can be viewed
as if we slightly move the contour from the real axis into the complex
plane $z=t+\i\varepsilon_m$ near the points $t_m$, where the momentum-space
loop $p(t)$ has discontinuities, by means of a harmonic function.} 
\be
\dot p(t)= \sum_m \Delta p_m 
\frac{\varepsilon_m}{\pi[\varepsilon_m^2+(t-t_m)^2]}\,,
\label{regdotpt}
\ee
which reproduces the sum of deltas as $\varepsilon_m\to0$.
The boundary value of the Liouville field now enters through  
\be
\varepsilon_m=\varepsilon \e^{-\varphi(t_m)/2}\,.
\label{27}
\ee

To obtain the scattering amplitudes, we
calculate $\int \d t \d t' \dot p(t) \dot p(t') \log|s(t)-s(t')|$ 
in the exponent with thus regularized $p(t)$:
\begin{eqnarray}
\lefteqn{\int \d t\, \d t' \, \dot p(t) \cdot \dot p(t') \ln|s(t)-s(t')|} 
\non &&=
\sum_{m,j}  \Delta p_m \cdot \Delta p_j \int
\frac{\varepsilon_m\d t }{\pi[\varepsilon_m^2+(t-t_m)^2]} \int
\frac{\varepsilon_j \d t'}{\pi[\varepsilon_j^2+(t'-t_j)^2]} \ln|s(t)-s(t')|.
\end{eqnarray}
For $j \neq m$ and $|t_m-t_j|\gg \varepsilon_m,\varepsilon_j$ 
the integral is the same as in the nonregularized case and gives
\be
\sum_{j\neq m}  \Delta p_m \cdot \Delta p_j \ln|t_m-t_j|, 
\label{no1}
\ee
while for $j=m$ we have
\be
\int 
\frac{\varepsilon_m\d t }{\pi[\varepsilon_m^2+(t-t_m)^2]}\int
\frac{\varepsilon_m \d t'}{\pi[\varepsilon_m^2+(t'-t_m)^2]} \ln|s(t)-s(t')|
= \ln \left[2\varepsilon_m \dot s(t_m)\right]+ {\cal O}(\varepsilon_m)\,.
\label{no2}
\ee
To complete the consideration, we note that
\be
\varepsilon_m  \dot s(t_m)= \varepsilon \e^{-\varphi(s_m)/2}
\ee
in view of Eqs.~\rf{115}, \rf{27}.

Adding \rf{no1} and \rf{no2} and using \eq{discLiousi}, we
reproduce the principal value prescription of Refs.~\cite{MO08,MO09}
that leads to the above off-shell scattering amplitudes,
provided \eq{discLiousi} is satisfied, where $s_{K_m}=t_m$ but $s_{K_m\pm1}$ 
live at the intermediate points as before.
Now these $s_{K_m\pm1}$'s are needed only to define the measure
for the integration over reparametrizations. 
We can alternative rewrite it via the Liouville field
and then not to introduce $s_i$'s at the intermediate points at all.
Those are needed, however, to manage the path integral over
reparametrizations.
 
\section{The shape of ``most important'' contours}

The ``most important'' contour \rf{1st} is given for thus smeared $p(t)$ by
\be
x_*(t)=\frac 1K p(t)= \frac 1K \sum_m \Delta p_m  \left(
\frac12 +\frac 1\pi \arctan \frac{(t-t_m)}{\varepsilon_m} \right)
=\frac 1{\pi K} \sum_m \Delta p_m  \arctan \frac{(t-t_m)}{\varepsilon_m}\,,
\label{amost}
\ee
which is nothing but a regularized  Eq.~(5) of \cite{MO09}.

For  the 4-particle case we express the four-vectors $\Delta p^\mu_m$
in the center-mass frame
through $E$, $p=|\vec p_m|$ and the scattering angle $\theta$ by
\bea
\Delta p_1 &=& \left( E,p,0,0 \right) \non
\Delta p_2 &=& \left( E,-p,0,0 \right) \non
\Delta p_3 &=& \left( -E,p \cos \theta ,p \sin \theta,0 \right) \non
\Delta p_4 &=& \left( -E,-p \cos \theta ,-p \sin \theta,0 \right) .
\label{centmass}
\eea

The scattering amplitude involves the integration over the variable
\be
x=\frac{(t_2-t_1)(t_4-t_3)}{(t_3-t_1)(t_4-t_2)},
\label{xfors2}
\ee 
which the integrand does depend on, as well as the integration over $t_1$,
$t_3$ and $t_4$, which factorizes. 
At large Mandelstam's variables $-s$ and $-t$ ($-t\la-s$)
the integral over $x$ is dominated by the saddle point
\be
x_*=\frac{s}{s+t},
\label{xsp}
\ee 
so it is fixed for given $s$ and $t$. On the contrary, the integral over 
the values of $t_4-t_1$ and $t_4-t_3$ gives a factor, which  equals 
the volume of the projective group. 
We can set, therefore, $t_1$, $t_3$ and $t_4$ equal to some values, 
keeping in mind that one
such contour belongs, in fact, to a whole family of contours,
which are equally important.

Choosing momenta in the 4-particle case according to \eq{centmass}
and using the projective invariance to set $t_1=0$, $t_3=1$, $t_4=\infty$,
that also fixes $t_2=x_*$ at large $s$ and $t$ according to Eqs.~\rf{xfors2} 
and \rf{xsp}, we arrive at the ``most important'' contours of the shape of
rectangles, which are depicted
for $t/s=-0.4$   
and $t/s=-0.04$ in Fig.~\ref{fi:amost04}   
and Fig.~\ref{fi:amost004}, respectively. They degenerate when $t/s\to0$.
\FIGURE{\vspace*{3mm}
\includegraphics[width=6cm]{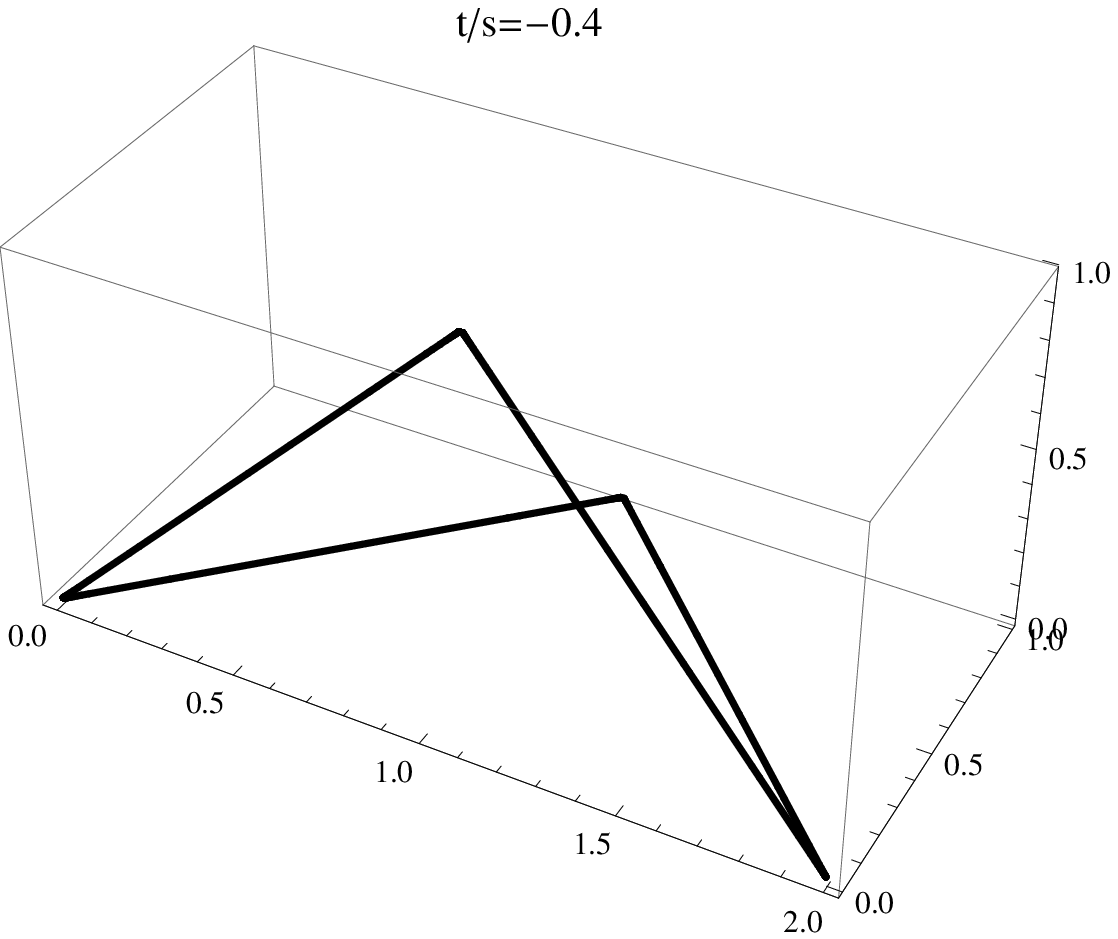} \hspace*{1cm}
\includegraphics[width=7cm]{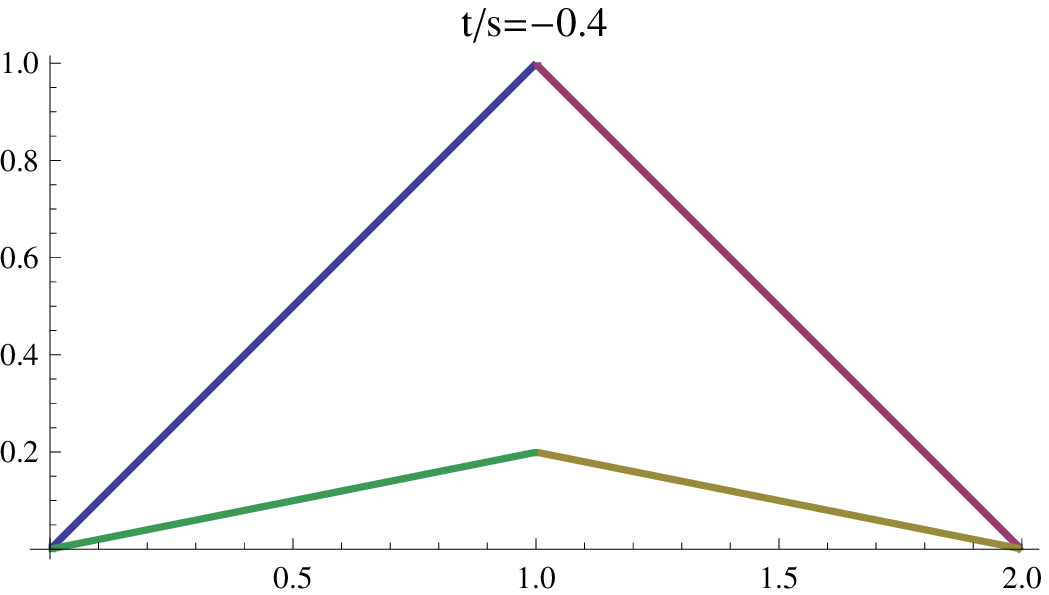}  
\caption[]{Typical loop $\C_*$ for $t/s=-0.4$.}   
\label{fi:amost04}  }
\FIGURE{\vspace*{3mm}
\includegraphics[width=6cm]{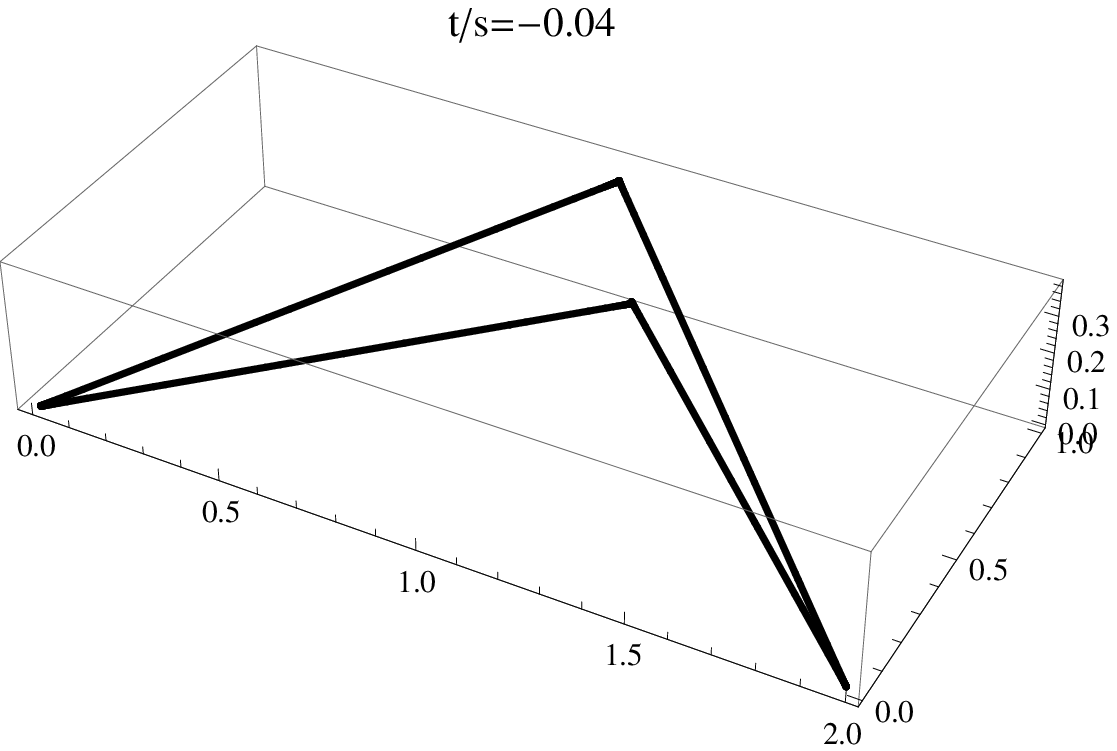} \hspace*{1cm}
\includegraphics[width=7cm]{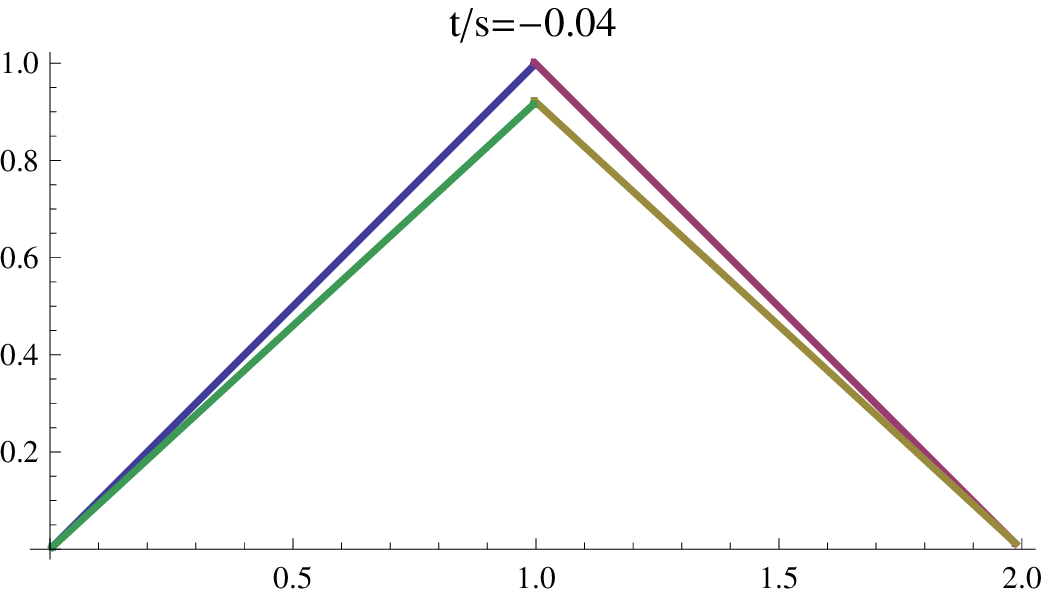}  
\caption[]{Typical loop $\C_*$ for $t/s=-0.04$.}   
\label{fi:amost004}  }

The minimal area 
\be
\K S_{\rm min}(C)=\alpha' t \log \frac st,
\ee
spanned by these contours,  
can be large even for small $t$. This will be enough to
justify the semiclassical approximation below. 
However, we have to have $|t|\ga \K$ for the transverse size of
the contour also to be large, 
which we be prescribed by a self-consistency of our approach in QCD.

\section{A generalization to $d$ dimensions\label{s:5}}

We are now in a position to perform the main task of this Paper:
to generalize the scattering amplitudes to $d<26$ dimensions.

For $d\neq 26$ the Liouville field does not decouple in the bulk, 
resulting in the additional factor to be inserted in the ansatz~\rf{ansatz}%
\footnote{The curvature term is not present for the upper half-plane
(UHP) parametrization.}
\be
W\left[x(\cdot)\right]=\int \D_{\rm diff} t \,\e^{-K S[x(t)]}\,
\hspace*{-5mm}\int\limits_{\varphi (x=s,y=0) = \varphi (s)} 
\hspace*{-5mm}\D \varphi \exp \left[
\frac{(d-26)}{48\pi} \int \d^2 z  
\left(\frac 12\partial_a \varphi \partial_a \varphi  +\mu \e^\varphi \right)
\right].
\label{Wfactor}
\ee
Here $z=x+i y\in\hbox{UHP}$ 
and $\varphi$ obeys the boundary condition
\be
\varphi (x=s,y=0) = \varphi (s)= 2 \ln \frac{\d t(s)}{\d s}
\label{bbcc}
\ee
at the real axis and we ignore the exponential $\e^{\varphi}$
in the Liouville action in \rf{Wfactor},
setting $\mu=0$. The second equality in \eq{bbcc} is valid again modulo
the scale mode $\varphi=\hbox{const.}$ which does not contribute to the 
Liouville action in \eq{Wfactor} when $\mu=0$. 

As was first shown in Refs.~\cite{FTs82,DOP84}, 
the presence of the additional factor 
in \rf{Wfactor} is crucial for reproducing the L\"uscher term,
describing semiclassical fluctuations of a string, for a rectangle
when $d\neq26$. An analogous result holds also for an ellipse, the structure
of the L\"uscher term for which is advocated in Ref.~\cite{MO10}.  

It is clear that the scattering amplitude which is a counterpart
of \eq{Wfactor} 
is given by \eq{ampl1} with the same additional factor
inserted in the integrand. The analysis of the path-integral in thus
obtained amplitude is beyond the scope of this paper.

We shall use instead a more simply tractable modification of 
the disk amplitude~\rf{ansatz},
which has been recently proposed by the authors~\cite{MO10}:
\begin{equation}
W\left[x(\cdot)\right]=\int \D_{\rm diff} t \,\e^{-K S[x(t)]}\,
\left(\det O\right)^{-(d-26)/48}
\label{modansatz}
\end{equation} 
with the Douglas integral $S[x(t)]$ given by \eq{Di} and
the determinant of the operator $O$ 
given by the Gaussian path integral
\be
\left(\det O\right)^{-1/2}= \int \D\beta(t) \e^{-K S_2[\beta]} 
\label{III2}
\ee
with the quadratic action 
\begin{equation}
S_2\left[\beta\right] = \frac 1{4\pi}  \int\d s_1 \pint \d s_2 
\frac{\dot x(t(s_1))\cdot \dot x(t(s_2))}{(s_1-s_2)^2 } 
\left[\beta(s_1)-\beta (s_2)\right]^2.
\label{AAA2}
\end{equation}
While the operator $O$ lives in the boundary, it captures semiclassical
transverse fluctuations of the minimal surface reproducing 
the L\"uscher term for a rectangle or an ellipse.
In $d=26$ \eq{modansatz} reduces to the Polyakov ansatz~\rf{ansatz}. 

It is important to emphasize that the additional determinant in 
Eq.~(\ref{modansatz}) does not effect the classical limit, which is
given by the exponential of the minimal area $ S_{\rm min}$, but contributes to
a pre-exponential. The parameter of the semiclassical expansion
is thereby $1/\K S_{\rm min}$, so for
large loops we can restrict ourselves with a semiclassical approximation. 

The functional Fourier transformation~\rf{msWL} 
of \rf{modansatz} is again calculable 
because the path integral over $x$ is Gaussian. 
In the semiclassical approximation we obtain again
\be
A\left[p(\cdot)\right]= W \left[\,x_*(\cdot)=p(\cdot)/K\,\right]\,,
\label{Psip}
\ee
where $W\left[x(\cdot)\right]$ is given by Eqs.~\rf{modansatz}, \rf{III2},
\rf{AAA2} with
$x(t)$  substituted by the ``most important'' contour~\rf{amost}.

\section{Semiclassical Regge trajectory}

Substituting the smeared step function \rf{amost} into \eq{AAA2} and
repeating the above calculation, we get
\bea
\lefteqn{K S_2 = \alpha' \sum_{i>j} \Delta p_i\cdot \Delta p_j
\int \frac{\d x}{\pi(1+x^2)}\int\frac{\d y}{\pi(1+y^2)}
\frac{\left[ \beta(t_i+\varepsilon_i x) 
- \beta(t_j+\varepsilon_j y)\right]^2}
{\left[s_*(t_i+\varepsilon_i x) -s_*(t_j+\varepsilon_j y)\right]^2}  }
\non &&\hspace*{1cm}+ 
\alpha' \sum_i\frac{\Delta p_i^2}2
\int \frac{\d x}{\pi(1+x^2)}\int\frac{\d y}{\pi(1+y^2)}
\frac{\left[ \beta(t_i+\varepsilon_i x) 
- \beta(t_i+\varepsilon_i y)\right]^2}
{\left[s_*(t_i+\varepsilon_i x) -s_*(t_i+\varepsilon_i y)\right]^2} \nonumber\\ 
&&\hspace*{4mm}\stackrel{\varepsilon\to0}\to 
\alpha' \sum_{m>n} \Delta p_m\cdot \Delta p_n
\frac{\left[ \beta(t_m) 
- \beta(t_n)\right]^2}
{\left[s_*(t_m) -s_*(t_n)\right]^2}
+\alpha'\sum_m\frac{ \Delta p_m^2}2
\frac{ \dot \beta^2(t_m) }
{\dot s^2_*(t_m)}.
\label{S333}
\eea

To treat the path integral over reparametrizations in \eq{III2}, 
we replace \rf{S333} by its discretized version
\begin{equation}
K S_2 = \alpha' \sum_{K_i>K_j} \Delta p_{K_i}\cdot \Delta p_{K_j}
\frac{(\beta_{K_i}  -\beta_{K_j})^2}
{(s_{K_i}-s_{K_j})^2} 
+ \alpha'\sum_{K_i}\frac{ \Delta p_{K_i}^2}2
\frac{\left(\beta_{K_{i}+1}-\beta_{K_{i}-1}\right)^2}
{\left(s_{K_{i}+1}-s_{K_{i}-1}\right)^2},
\label{S334}
\end{equation}
where the indices are labeled as before. The first term on the right-hand side
of \eq{S334} [or \eq{S333}]
vanishes since $\beta_{K_i}=0$ because of $s_*(t_m)=t_m$. 
We have also verified that the obtained result does not change for a wide
class of discretization, in particular, for another discretization of 
$\dot\beta(t_m)$.

The integral over $\beta_{i}$'s 
at the intermediate points now decouples from the one over $\beta_{K_{i}\pm1}$. 
This shows the difference between the path integrals over reparametrizations
around the minimizing trajectory $s_*(t)$ for a smooth contour like an ellipse 
or rectangle, which are considered in Ref.~\cite{MO10}, and the
smeared step function~\rf{amost}. In the former case we have a Gaussian
integral that reproduces the L\"uscher term for large $K S_{\rm min}$,
while in the latter case the quadratic action vanishes and we deal with
zero modes. The reason is that the only restriction on the minimizing
function $s_*(t)$ for a step-wise contour is $s_*(t_m)=t_m$, as was pointed
out in Ref.~\cite{MO09}, so $s_*(t)$ is arbitrary for $t_{m-1}<t<t_m$. 

If $\Delta p_m^2=0$ all $\beta_i$'s can be treated on equal footing
as zero modes. 
In this case the quadratic approximation is not applicable and
we have to integrate over reparametrizations with the whole measure
$\D_{\rm diff} s$ given by \eq{altmeasure}. We write, therefore,
\begin{eqnarray}
\det\, O^{-1/2} &= &\prod_m \int\limits_{t_{m-1}<t_m} 
\d t_m \,\frac{(s_{K_m+1}-s_{K_m-1})}{(s_{K_m+1}-t_{m})(t_{m}-s_{K_m-1})}\, \non
&& \hspace*{1cm}\times
\d s_{K_m-1} \,\frac{(t_{m}-s_{K_m-2})}{(t_{m}-s_{K_m-1})(s_{K_m-1}-s_{K_m-2})}\, \non
&& \hspace*{1cm}\times
\d s_{K_m-2} \,\frac{(s_{K_m-1}-s_{K_m-3})}{(s_{K_m-1}-s_{K_m-2})(s_{K_m-2}-s_{K_m-3})}\,
\cdots \nonumber \\
&& \hspace*{1cm}\times\d s_{K_{m-1}+2} \,\frac{(s_{K_{m-1}+3}-s_{K_{m-1}+1})}
{(s_{K_{m-1}+3}-s_{K_{m-1}+2})(s_{K_{m-1}+2}-s_{K_{m-1}+1})}\, \non && \hspace*{1cm}\times
\d s_{K_{m-1}+1} \,\frac{(s_{K_{m-1}+2}-t_{m-1})}
{(s_{K_{m-1}+2}-s_{K_{m-1}+1})(s_{K_{m-1}+1}-t_{m-1})}\,.
\label{meee}
\end{eqnarray}

The multiple integral in \eq{meee} coincides with the one considered in
Sect.~\ref{s:3} if we set there $\Delta p_i=0$. 
The resulting formula for the integral is of the type of 
the contribution from the measure to \eq{A15p} and reads
\begin{equation}
\det\, O^{-1/2} \propto (1-x)^{-1}\sim \alpha' s \,.
\label{dett}
\end{equation}
The same result can be obtained also for $\Delta p_i^2\neq 0$, substituting
$ s_{i}=s_*{}_{i}+\beta_{i} $
in \eq{meee}.

The determinant~\rf{dett} results in an additional factor of $s^{(d-26)/24}$ 
in the amplitude 
and there are no other such factors because
typical values of $(u_m-v_m)$ in \eq{calKD} are not small.
We therefore obtain the following value of the Regge intercept %
\begin{equation}
\alpha(0)=\alpha_0 + \frac{(d-26)}{24}\,,
\end{equation}
where $\alpha_0=1$ is the value calculated above in $d=26$. 

Finally we get in $d=4$
\be
\alpha(0)=\frac{(d-2)}{24}\approx 0.083 \,,
\label{inter}
\ee 
which is to be compared with the intercept of a quark-antiquark 
Regge trajectory.
The value~\rf{inter} is smaller than the experimental value of 
$\alpha(0)\approx 0.5$ for the $\rho-A_2-f$ meson Regge trajectory.
A possible explanation of this discrepancy is that  
the chiral symmetry apparently is not yet spontaneously broken 
in our consideration as is clear from the fact that the amplitude does not
have Adler's zero (i.e.\  does not vanish with vanishing $s$ and $t$
in the chiral limit).
Then the breaking of the chiral symmetry may shift the intercept up,
like this happens~\cite{Lov68} in the Lovelace--Shapiro dual models.

The reason why we consider such a quark-antiquark Regge trajectory is because
we are dealing with the disk amplitude, which is associated with 
planar diagrams and correspondingly a quark-antiquark Regge trajectory 
in large-$N$ QCD, as we shall now discuss.

\section{Application to QCD}

As is already mentioned, the disk amplitude~\rf{modansatz} describes
the asymptote of Wilson loops in large-$N$ QCD.
 
$M$-particle scattering amplitudes in large-$N$ QCD are expressed
through the Green functions of $M$ colorless composite quark
operators (e.g.\ ${\bar q}(x_i) q(x_i)$) in terms of the sum
over all Wilson loops passing via the points $x_i$ ($i=1,\ldots,M$), where the
operators are inserted.
The on-shell $M$-particle scattering amplitudes can be obtained from these
Green functions by the standard Lehman--Symanzik--Zimmerman reduction.
Representing $M$ momenta of the (all incoming) particles by the
differences
$ %
\Delta p_i= p_{i\!-\!1}-p_i
$ 
and introducing a piecewise constant momentum-space loop
\begin{equation}
p(t)=p_i \qquad \hbox{for}~ t_i<t<t_{i\!+\!1} \,,
\label{piecewise}
\end{equation}
we obtain
\begin{eqnarray}
A\left(\Delta p_1,\ldots, \Delta p_M \right) &\propto&
\int\nolimits_0^\infty \d \Tau\, \Tau^{M-1} \e^{-m \Tau} 
\int\nolimits_{-\infty}^{+\infty} \frac{\d t_{M-1}}{1+t_{M-1}^2}
\prod_{i=1}^{M\!-\!2}\int\nolimits_{-\infty}^{t_{i\!+\!1}} \frac{\d t_i}{1+t_i^2} 
\non &&
\times  \hspace*{-6mm}
\int\limits_{z(-\infty)=z(+\infty)=0} \hspace*{-6mm} \D z(t) \,
\e^{\i \int \d \tau\, \dot z (t)\cdot p(t)}\,
J[z(t)] \, W[z(t)].  
\label{117}
\end{eqnarray}
Here we do not integrate over $z(-\infty)=z(+\infty)$, which would produce 
the (infinite) volume factor because of translational invariance. 

For spinor quarks and scalar operators the weight for the path integration 
in \eq{117} is
\be
J[z(t)]=\int \D k(t)\;{\rm sp\;} {\boldmath P}
\e^{\i \int \d t\,
[\dot z (t)\cdot k(t)- \Tau \gamma\cdot k(t)/(1+t^2)]} ,
\label{J}
\ee
where sp and the path-ordering refer to $\gamma$-matrices.
In \eq{117} $W(C)$ is the Wilson loop in pure Yang--Mills theory at large $N$
(or quenched), $m$ is the quark mass and $\Tau$ is the proper time.
For finite $N$, correlators of several Wilson loops have to be taken into 
account.

Substituting the ansatz~\rf{ansatz} into \eq{117}, we can perform the
Gaussian integral over $z(t)$ to obtain
\begin{eqnarray}
A\left(\Delta p_1,\ldots, \Delta p_M \right) &\propto&
\int\nolimits_0^\infty \d \Tau\, \Tau^{M-1} \e^{-m \Tau} 
\int\nolimits_{-\infty}^{+\infty} \frac{\d t_{M-1}}{1+t_{M-1}^2}
\prod_{i=1}^{M\!-\!2}\int\nolimits_{-\infty}^{t_{i\!+\!1}} \frac{\d t_i}{1+t_i^2}\non 
&& \times \int \D k(t)\;{\rm sp\;} {\boldmath P}
\e^{-\i\Tau  \int \d t\,\gamma\cdot k(t)/(1+t^2)} \,
W\Big[x_*(t)=\frac1K \left(p(t)+k(t)\right)  \Big]. \non &&
\label{118}
\end{eqnarray}

As distinct from its stringy counterpart~\rf{Psip},
the right-hand side of \eq{118} has the additional path integration over $k(t)$,
which emerges from Feynman's disentangling of the $\gamma$-matrices.
But for the case, where $m$ 
is small and/or $M$ is very large, the integral over $\Tau$ in \eq{117} 
is dominated by large $\Tau\sim (M-1)/m$. 
Noting that typical
values of $k\sim 1/\Tau$ are essential in the path integral over $k$
for large $\Tau$, we can disregard $k(t)$ in the argument of $W$ in \eq{118},
so the path integral over $k$ factorizes. 
We finally obtain~\cite{MO08} from \eq{117} the product of the momentum-space
disk amplitude $A\left[p(t)\right]$ times factors which do not depend on $p$.
The substitution of the ansatz~\rf{modansatz} into \eq{117} for $d<26$ 
results in a more complicated path integral over $z(t)$ which is, however,
Gaussian in the semiclassical approximation, reproducing again \eq{118}.

Thus for the ansatz~\rf{modansatz} $A\left[p(t)\right]$ 
coincides with $W\left[x_*(t)\right]$, where $x_*(t)$ is given by \eq{amost}
and the path integral over reparametrizations in \eq{modansatz} goes over 
the functions $s(t)$, obeying $s(t_m)=t_m$. Denoting 
\be
\frac{\d t}{\d s} =\e^{\varphi(s)/2} \,,
\label{deno}
\ee 
we can finally rewrite \eq{Psip} in the form displayed in \eq{ampl1}. 
Therefore, \eq{118} exactly reproduces for piecewise constant $p(t)$
the (off-shell) amplitude~\rf{amplitudep22} as 
$m\rightarrow0$ and/or $M\to\infty$!
We thus conclude that the quark-antiquark Regge trajectory in large-$N$
QCD is linear in the semiclassical approximation with the intercept given 
by \eq{inter}.

\section{Separation of pQCD and npQCD}

In QCD string is stretched between quarks, when they are separated by
large distances. Its emergence is due to nonperturbative effects.
Alternatively, perturbative QCD (pQCD) works at small distances,
where the reggeization of $\bar q q$ is due to double logarithms
in quark amplitudes, 
as was first pointed out by Kirschner and Lipatov~\cite{KL83} 
and further investigated in Refs.~\cite{Brodsky,BL03}. 
They found for scattering amplitudes with an exchange of a
quark-antiquark pair:
\begin{equation}
\hbox{{pQCD double logs}}~ 
\propto~ I_1 \left(\omega \ln \frac{|s|}{\mu^2}
\right),
\quad \quad \omega=\sqrt{\frac{{g}^2(t) C_F}{2\pi^2}}\,, 
\label{pQCDl}
\end{equation}
where $\mu\sim 1~{\rm GeV}$ is an IR cutoff (usually given in 
pQCD by a transverse mass). In spite of \eq{pQCDl} is obtained with
the double logarithmic accuracy, the coupling
${g}^2 $ is replaced by ${g}^2 (t)$ because of the 
anticipated charge renormalization.
We then have asymptotically the Regge behavior
\be
\hbox{pQCD double logs}~ \propto~ \left( \frac{s}{\mu^2} 
\right)^{\omega(t)}
\label{ReggepQCD}
\ee
with almost constant $\alpha(t)\approx 0.25\div 0.5$ for
$t\la -1~{\rm GeV}^2$, where pQCD applies.

Experimental data for $\alpha(t)$ extracted for $t<0$
from inclusive $\pi^0$ production in $\pi^-p$ collisions
and from the exclusive process $\pi^-p\to \pi^0 N$ 
are analyzed, respectively, in Refs.~\cite{Brodsky,Kai06}.
The data for $t>-2~{\rm GeV}^2$ as well as the spectrum of resonances for $t>0$ 
are well described by a linear $\rho$-meson Regge trajectory with the intercept 
$\alpha(0)\approx 0.5$.
Alternatively, the dependence~\rf{ReggepQCD} for $\alpha(t)$ would be 
represented by an almost horizontal line for \mbox{$t\la -1~{\rm  GeV}^2$}. 
As is emphasized in Ref.~\cite{Kai06}, 
the scattering data do not fit such a behavior of $\alpha(t)$
in pQCD.

In \eq{117}, which relates scattering amplitudes and the Wilson loops,
pQCD resides in the contribution from small loops, while nonperturbative
stringy effects reside in the contribution of large loops, where
the ansatz~\rf{modansatz} is applicable. 
The total amplitude is obviously a sum of these two contributions,
which may be separated by splitting the integral over $\Tau$
into two regions: $\Tau<\tau_{\rm max}$ and  $\Tau>\tau_{\rm max}$
with $\tau_{\rm max}\sim K$.
We associate them with pQCD and npQCD (nonperturbative QCD), respectively.%
\footnote{Sometimes the terms ``hard'' and ``soft'' are also used,
respectively, for the pQCD and npQCD contributions to Regge trajectories.}
For the pQCD region $\tau_{\rm max}$ plays the role of
an IR cutoff, as $1/\mu^2$ does in Minkowski space.  
If nonperturbative effects have not been taken into account,
the pQCD diagrams would be cut in infrared at much smaller
momenta of the order of the quark mass $m$.

Because the total amplitude is the sum of both pQCD and npQCD contributions
which decrease with $t$ as different powers of $s$,
the relative coefficient is of most importance at large but finite $s$.
The npQCD contribution to \eq{117} has a relative factor of 
$\left(\sqrt{\K}/m\right)^M$, which is large for small $m$.
It is worth analyzing whether or not this would be numerically enough to
explain the dominance of the npQCD contribution for $t>-2~{\rm GeV}^2$.

\section{Conclusion} 

We have shown in this Paper how to deal with the path integral
over reparametrizations of the boundary of an open string (or, equivalently,
over the boundary value of the Liouville field in the Polyakov formulation). 
This path integration over reparametrizations is crucial to obtain
consistent off-shell scattering amplitudes in the critical dimension $d=26$.

We used the representation of the disk amplitude as the path integral
over reparametrizations of the (exponential of the) boundary action, known
as the Douglas integral, which is quadratic in the embedding-space
coordinates and the functional Fourier transformation
can be performed for this reason. Therefore, 
it was crucial for the success of calculations that all path integrals are
Gaussian except for the one over reparametrizations, which 
can be partially done and the remainder
reduces to an integration over the Koba--Nielsen variables.

For $d<26$ we used a similar representation in the form of a
boundary path integral, which is equivalent 
in the semiclassical approximation to the
path integral over the Liouville field in the bulk, that accounts
for transverse fluctuations of the minimal surface.
We have obtained from it, again in the semiclassical approximation,
a Regge-behaved scattering amplitude with a linear Regge trajectory 
of the intercept $(d-2)/24$, that remarkably coincides with the 
one obtained from the spectrum~\cite{Arv83} of the long Nambu--Goto string,
which is asymptotically consistent~\cite{p85} even for $d\neq 26$.
This apparently illustrates its equivalence to the Polyakov string in the
semiclassical approximation.

An application of thus obtained scattering amplitudes of noncritical
string theory is the theory of strong interaction of hadrons. 
In our case a vector excitation  is
associated with a massive $\rho$-meson rather than a massless gauge field.
We have demonstrated how this result applies to the quark-antiquark Regge
trajectory in QCD, where the boundary action describes 
asymptotic behavior of Wilson loops of large size.

A natural question is as to what are the approximations made in the derivation? 
Remarkable, there are surprisingly few of them. We have to have the limit
of large number of colors to justify the quenched approximation.
Also small quark mass and/or the large number of external legs of
the amplitude essentially simplify the result --- otherwise the
amplitude involves a rather complicated path integral associated with
spin degrees of freedom. The amplitude applies in the Regge kinematical
regime of asymptotically large $s$ and fixed $t$,
associated with small scattering angle or fixed momentum transfer, 
where the nonperturbative stringy behavior of
the Wilson loop dominate over perturbative QCD. From experiment we expect
this should be the case for $t>- 2 ~\hbox{GeV}^2$, while a future
investigation of the relative strength of perturbative and nonperturbative
effects is required to justify this domain. For much larger values of 
$-t\sim s$ there are no longer reasons to expect the nonperturbative stringy  
effects to dominate over perturbation theory. On the contrary,
perturbative QCD should dominate for large $-t$, so 
the exponential stringy behavior of the differential cross-section 
is expected to change for a field-theoretical power-like behavior
of perturbative QCD at small distances.

In the region, where the area-law behavior of the Wilson loop sets in,
we expect large loops to dominate the sum over path on the
right-hand side of \eq{117}, so that the semiclassical approximation 
we used to find the Regge intercept is applicable.
Therefore, our semiclassical approximation is justified by the Regge
kinematics.
A phenomenological argument in favor of this picture is that
the Regge trajectory is linear in the semiclassical approximation
and higher orders would most probably lead to 
a bending of the Regge trajectory, which is not seen in experiment for the 
quark-antiquark Regge trajectory.

Our final comment concerns an ambiguity of the scattering amplitudes
which is due to different discretizations of the measure in the path
integral over reparametrizations. In Ref.~\cite{MO09} we used
the discretization~\rf{PRDmeasure} which resulted in the zero intercept,
while the discretization~\rf{altmeasure} of this Paper
has resulted in the unit intercept in $d=26$. We have also verified
these two numbers do not change, when next to neighbor points are
involved in the discretization of the measure.
These are apparently two universality classes.
It is worth further investigating of this issue, in particular, of
the meaning of \eq{altconvolution} from the point of view of an associated 
stochastic process, like it was done in Ref.~\cite{BM09} for the 
measure~\rf{PRDmeasure}.

\begin{acknowledgments}
We are indebted to Alexander Gorsky, Alexei Kaidalov, Gregory Korchemsky, 
Andrei Mironov, Niels Obers, and Konstantin Zarembo 
for encouragement and useful discussions.
\end{acknowledgments}

\appendix

\section{Integration with alternative measure (3.2)\label{appC}}

In this appendix we derive \eq{altconvolution} which plays a crucial role
in the integration over reparametrizations with the measure~\rf{altmeasure}.

Let us represent
\be
\left[\frac{(s_{i+1}-s_{i-1})}{(s_{i+1}-s_{i})(s_{i}-s_{i-1})}\right]=
\left[\frac{1}{(s_{i+1}-s_{i})}+\frac1{(s_{i}-s_{i-1})}\right].
\ee
Multiplying three brackets, we get on the left-hand side of \eq{altconvolution}
eight integrals of the type
\be
\int_{s_1}^{s_3} \d s_2\, s_{32}^{a-1} s_{21}^{b-1} =
s_{31}^{a+b-1} B\left(a,b\right)
\ee
with
\be
 B\left(a,b\right)=\int_{1}^\infty \d y \,
\frac{\left(y^{a-1}+y^{b-1}\right)}{(1+y)^{a+b}}\,.
\label{beta}
\ee
When the integral in \eq{beta} is convergent, it is of course just 
the beta-function, 
but a regularization is required for $a,b= 0,-1$ as in our case.
Integrating over $s_2$, we obtain
\bea
\lefteqn{ \int_{s_1}^{s_3} \d s_2 
\left[\frac{s_{42}}{s_{43} s_{32}}\right]
\left\{
\left[\frac{s_{31}}{s_{32} s_{21}}\right] 
\left[\frac{s_{20}}{s_{21} s_{10}}  \right] +c\frac{1}{s_{21}^2} 
\right\} }
\non
& &=\left[ B\left(0,0\right)+B\left(1,-1\right)\right]s_{43}^{-1} s_{31}^{-1} 
+
 \left[ B\left(0,1\right)+B\left(1,0\right)\right]s_{43}^{-1} s_{10}^{-1}
\non &&~+
 \left[ B\left(-1,0\right)+B\left(0,-1\right)\right] s_{31}^{-2}
+
 \left[ B\left(0,0\right)+B\left(-1,1\right)\right] s_{31}^{-1} s_{10}^{-1}
\non &&~+c\left[ B\left(0,-1\right) s_{31}^{-2}+
 B\left(1,-1\right) s_{43}^{-1}s_{31}^{-1}\right]
 \,.
\label{altintegra}
\eea
The four coefficients read explicitly as
\be
 B\left(0,0\right)+B\left(1,-1\right)= 
B\left(0,0\right)+B\left(-1,1\right)=
B\left(-1,0\right)=B\left(0,-1\right)
=\int_1^\infty \d y\, \left( 1+\frac 1y \right)^2
\label{ee1}
\ee
and
\be
B\left(0,1\right)+B\left(1,0\right)
=2\int_1^\infty \frac{\d y}y \,.
\label{ee2}
\ee
The integral over $y$ in \eq{ee2} is logarithmically divergent at 
large $y$ and can be regularized, e.g.\ by
\be
\int_1^\infty \frac{\d y}{y^{1+\delta}}=\frac1\delta \,.
\label{zeta1}
\ee
This type of an analytic regularization simultaneously regularizes
\be
\int_1^\infty \frac{\d y}{y^{\delta}}=\frac1{\delta-1} 
\label{zeta2}
\ee
in \eq{ee1} making the difference between the two to be finite.
It preserves, therefore, the projective invariance of the 
integral~\rf{altintegra} as $\delta\to0$.

We thus find
\bea
\lefteqn{\int_{s_1}^{s_3} \d s_2 
\left[\frac{s_{42}}{s_{43} s_{32}}\right]
\left\{
\left[\frac{s_{31}}{s_{32} s_{21}}\right] 
\left[\frac{s_{20}}{s_{21} s_{10}}  \right] +c\frac{1}{s_{21}^2} 
\right\}}\non &&
=\frac 1\delta 
\left\{\left[\frac{s_{41}}{s_{43} s_{31}}  \right]
\left[\frac{s_{30}}{s_{31} s_{10}}  \right]+(c+1)\frac{1}{s_{31}^2} \right\}
+{\cal O}\left(\delta^0\right),
\label{altconvolution1}
\eea
where both terms are manifestly projective-invariant, which proves 
\eq{altconvolution}.

We finally note that the formulas \rf{zeta1}, \rf{zeta2} are of the type of
the ones for the regularization via the $\zeta$-function
\be
\sum_{1}^\infty \frac{1}{n^{1+\delta}}=\zeta(1+\delta)\,,
\ee
since
\be
\zeta(1+\delta) = \frac{1}{\delta}+{\cal O}\left(\delta^0\right).
\ee
This is because we can substitute the integration in \eq{zeta1} by
a summation, which does not change the divergent part of the integral.
Analogously 
\be
\zeta(0)=-\frac12
\ee
is the counterpart of \eq{zeta2}.

The $n$-fold integral over reparametrizations at the intermediate points
can be calculated applying \eq{altconvolution} $n$ times to give
\bea
 \lefteqn{\prod_{i=1}^n \int_{s_1}^{s_{i+2}}\d s_{i+1}\, 
\left[\frac{(s_{n+3}-s_{n+1})}{(s_{n+3}-s_{n+2})(s_{n+2}-s_{n+1})}  \right]
 \prod_{i=1}^n 
\left[\frac{(s_{i+2}-s_{i})}{(s_{i+2}-s_{i+1})(s_{i+1}-s_{i})}  \right]   }
\non && \hspace*{4cm}\times 
\left[\frac{(s_{2}-s_{0})}{(s_{2}-s_{1})(s_{1}-s_{0})} \right] \non &&
  \propto \left\{
\left[\frac{(s_{n+3}-s_1)}{(s_{n+3}-s_{n+2})(s_{n+2}-s_{1})} \right] 
\left[\frac{(s_{n+2}-s_0)}{(s_{n+2}-s_{1})(s_{1}-s_{0})} \right] 
+n \frac{1}{(s_{n+2}-s_1)^2} \right\} \,.
\label{n-fold}
\eea
We denote the right-hand side of \eq{n-fold} as 
\bea
D\left(s_{n+3},s_{n+2} ; s_1, s_0  \right)&=&
\frac{1}{(s_{n+3}-s_{n+2})(s_{n+2}-s_{1})}+
\frac{1}{(s_{n+3}-s_{n+2})(s_{1}-s_{0})} \non &&+
\frac{1}{(s_{n+2}-s_{1})(s_{1}-s_{0})} +
(n+1)\frac{1}{(s_{n+2}-s_1)^2} \,.
\label{defD}
\eea
For smooth functions $s(t)$ only the second term on the right-hand side
of \eq{defD} would be essential, when $n$ is large.
Alternatively, for discontinuous trajectories of the type discussed in
Ref.~\cite{BM09} the fourth term is expected to dominate because of
the large factor of $n$.

With thus defined $D$ we can write the amplitude 
in the form~\rf{amplitudep22} with ${\cal K}\left(\{t_m \}\right)$ given
by \eq{calKD}. In spite of the complicated structure of 
${\cal K}\left(\{t_m \}\right)$, the Regge behavior of 
the amplitude~\rf{amplitudep22} can be analyzed and is given by \eq{A15p}
for the general expression~\rf{defD}.



\begin{thebibliography}{99}


\bibitem{Pol87}
A.~M.~Polyakov,
{\it Quantum geometry of bosonic strings,}
  Phys.\ Lett.\  B {\bf 103}, 207 (1981);
{\it Gauge fields and strings},  
(Harwood Acad.\ Pub., Chur, 1987). 

\bibitem{ADN86}
  H.~Aoyama, A.~Dhar and M.~A.~Namazie,
  {\it Covariant amplitudes in Polyakov's string theory,}
  Nucl.\ Phys.\  B {\bf 267}, 605 (1986).


\bibitem{MO08}
  Y.~Makeenko and P.~Olesen,
{\it Implementation of the duality between Wilson loops and scattering
amplitudes in QCD},
  Phys.\ Rev.\ Lett.\  {\bf 102},  071602 (2009)  [arXiv:0810.4778 [hep-th]].

\bibitem{MO09}
  Y.~Makeenko and P.~Olesen,
{\it Wilson loops and QCD/string scattering amplitudes},
Phys.\ Rev.\ D  {\bf 80}, 026002  (2009)  [arXiv:0903.4114 [hep-th]].

\bibitem{BM09}
  P.~Buividovich and Y.~Makeenko,
{\it Path integral over reparametrizations: Levy flights versus random walks,}
Nucl.\ Phys.\ B {\bf 834} 453, (2010) [arXiv:0911.1083 [hep-th]].

\bibitem{MO10}
  Y.~Makeenko and P.~Olesen,
{\it Quantum corrections from a path integral over reparametrizations,}
Phys.\ Rev.\ D {\bf 82} v.4, (2010) [arXiv:1002.0055 [hep-th]].

\bibitem{Lov70}
C. Lovelace, {\it Simple $N$-Reggeon vertex}, 
Phys.\ Lett.\ B {\bf 32}, 490 (1970).

\bibitem{DiV92}
  P.~Di Vecchia,
{\it Multiloop amplitudes in string theories,} 
in  {\it String Quantum Gravity and Physics at the Planck Energy Scale,  
Erice 1992},  ed.\ N. Sanchez, (World Scientific, 1993), p.~16; \\
L.~Cappiello, A.~Liccardo, R.~Pettorino, F.~Pezzella, and R.~Marotta,
{\it Prescriptions for off-shell bosonic string amplitudes,}
  Lect.\ Notes Phys.\  {\bf 525},  466 (1999) [{arXiv:hep-th/9812152}];\\
A.\ Liccardo, F.\ Pezzella, and R.\ Marotta, 
{\it Consistent off-shell tree string amplitudes}, 
Mod.\ Phys.\ Lett.\  A {\bf 14}, 799 (1999) [{arXiv:hep-th/9903027}].

\bibitem{Arv83}
J.~F.~Arvis, {\it The exact $\bar qq$ potential in Nambu string theory},
Phys.\ Lett.\ B {\bf 127}, 106 (1983).

\bibitem{p85} 
P. Olesen, {\it Strings and QCD}, Phys.\ Lett.\ B {\bf 160}, 144 (1985).

\bibitem{Dou31}
J.~Douglas, {\it Solution of the problem of Plateau},
Trans.\ Am.\ Math.\ Soc. {\bf 33},  263 (1931).

\bibitem{Pol97}
A.~M.~Polyakov, Talk at the Workshop ``Particles, Fields and Strings'',
Vancouver, July 1997, unpublished.

\bibitem{Ryc02}
V.~S. Rychkov,  
{\it Wilson loops, D-branes, and reparametrization path integrals},
JHEP {\bf 0212}, 068 (2002)
[{arXiv:hep-th/0204250}].

\bibitem{AM07a}
  L.~F.~Alday and J.~Maldacena,
{\it Gluon scattering amplitudes at strong coupling,}
  JHEP {\bf 0706},  064 (2007) [arXiv:0705.0303 [hep-th]].

\bibitem{DSK07}
J.~M.~Drummond, G.~P.~Korchemsky, and E.~Sokatchev,
{\it Conformal properties of four-gluon planar amplitudes and Wilson loops,}
 Nucl.\ Phys.\ B {\bf 795}, 385 (2008) [arXiv:0707.0243 [hep-th]]; \\
  A.~Brandhuber, P.~Heslop, and G.~Travaglini,
{\it MHV Amplitudes in N=4 super Yang-Mills and Wilson loops,}
  Nucl.\ Phys.\  B {\bf 794},  231 (2008) [arXiv:0707.1153 [hep-th]]; \\
 J.~M.~Drummond, J.~Henn, G.~P.~Korchemsky, and E.~Sokatchev,
{\it On planar gluon amplitudes/Wilson loops duality,}
Nucl.\ Phys. B {\bf 795}, 52 (2008) [arXiv:0709.2368 [hep-th]].

\bibitem{AR08}
  L.~F.~Alday and R.~Roiban,
{\it Scattering amplitudes, Wilson loops and the string/gauge theory
 correspondence,} 
Phys.\ Rep.\ {\bf 468}, 153 (2008) [arXiv:0807.1889 [hep-th]].

\bibitem{FTs82}
E.~S.~Fradkin and A.~A.~Tseytlin,
{\it On quantized string models,}
Ann.\ Phys.\  {\bf 143}, 413 (1982).  

\bibitem{DOP84}
B. Durhuus, P. Olesen, and J.L. Petersen, 
{\it On the static potential in Polyakov's theory of the quantized string},
Nucl.\ Phys.\ B {\bf 232},  291 (1984).

\bibitem{Lov68}
  C.~Lovelace,
  {\it A novel application of Regge trajectories,}
  Phys.\ Lett.\  B {\bf 28}, 264 (1968); \\
J.~A.~Shapiro, {\it Narrow-resonance model with Regge behavior for
$\pi\pi$ scattering}, Phys.\ Rev.\ {\bf 179}, 1345 (1969).

\bibitem{KL83}
R.~Kirschner and L.~N.~Lipatov, 
{\it Double logarithmic asymptotics of quark scattering amplitudes 
with flavor exchange},
Phys.\ Rev.\ D {\bf 26}, 1202 (1982); 
{\it Double logarithmic asymptotics and Regge singularities of 
quark amplitudes with flavor exchange},
Nucl.\ Phys.\ B {\bf 213}, 122 (1983).

\bibitem{Brodsky} 
  S.~J.~Brodsky, W.~K.~Tang, and C.~B.~Thorn,
  {\it The reggeon trajectory in exclusive and inclusive large momentum transfer
  reactions,}
  Phys.\ Lett.\  B {\bf 318}, 203 (1993).

\bibitem{BL03}
  J.~Bartels and M.~Lublinsky,
{\it Quark antiquark exchange in $\gamma^* \gamma^*$ scattering,}
JHEP {\bf 0309}, 076 (2003) [arXiv:hep-ph/0308181].

\bibitem{Kai06}
A. B. Kaidalov, 
{\it Some problems of diffraction at high energies}, arXiv:hep-th/0612358.

\end{thebibliography}
\end{document}